# DNA cyclization and looping in the wormlike limit: normal modes and the validity of the harmonic approximation

Stefan M. Giovan,[1)] Andreas Hanke,[4)] and Stephen D. Levene[1,2,3,a)]

*Departments of [1]Molecular and Cell Biology, [2]Bioengineering, and [3]Physics, University of Texas at Dallas, Richardson, TX 75083 USA*

[4]*Department of Physics and Astronomy, University of Texas at Brownsville, Brownsville, TX 78520 USA*

For much of the last three decades Monte Carlo-simulation methods have been the standard approach for accurately calculating the cyclization probability, *J*, or J factor, for DNA models having sequence-dependent bends or inhomogeneous bending flexibility. Within the last ten years, however, approaches based on harmonic analysis of semi-flexible polymer models have been introduced, which offer much greater computational efficiency than Monte Carlo techniques. These methods consider the ensemble of molecular conformations in terms of harmonic fluctuations about a well-defined elastic-energy minimum. However, the harmonic approximation is only applicable for small systems, because the accessible conformation space of larger systems is increasingly dominated by anharmonic contributions. In the case of computed values of the J factor, deviations of the harmonic approximation from the exact value of *J* as a function of DNA length have not been characterized. Using a recent, numerically exact method that accounts for both anharmonic and harmonic contributions to *J* for wormlike chains of arbitrary size, we report here the apparent error that results from neglecting anharmonic behavior. For wormlike chains having contour lengths less than four times the persistence length the error in *J* arising from the harmonic approximation is generally small, amounting to free energies less than the thermal energy, $k_BT$. For larger systems, however, the deviations between harmonic and exact *J* values increase approximately linearly with size.

---
[a)] Author to whom correspondence should be addressed. E-mail: sdlevene@utdallas.edu.

## I. INTRODUCTION

We recount in this section some key developments in the theoretical treatment of DNA cyclization since the early 1980s, emphasizing the role that Don Crothers and his collaborators played in advancing this field. Although the Crothers group also contributed greatly to the experimental DNA-cyclization literature (sometimes in papers that combined experiments and theory), we elected not to review those important contributions here in order to avoid a lengthy digression.

Rigorous experimental measurements of the cyclization free energy of DNA molecules have been carried out since the ground-breaking work of Shore *et al.* [1] and Shore and Baldwin.[2,3] At that time experiments were well ahead of available theory for the dependence of the cyclization probability on DNA size and helical phase. The lack of a useful theoretical framework was due to the inherently challenging problem of modeling a stiff polymer chain having chain-end constraints in a way that also accounts for thermal fluctuations in such systems. The treatment of conformational fluctuations is important because they are significant even for DNA molecules with contour lengths substantially less than the polymer's persistence length.[4,5]

A major theoretical advance came from the work of Shimada and Yamakawa,[6,7] who developed a semi-numerical approach based on a series approximation to the cyclization probability of a uniform helical wormlike chain. The rigor and numerical accuracy of this theory permitted values of the persistence length, torsional rigidity, and helical repeat to be extracted from Shore and Baldwin's experimental data with confidence. However, there was a significant limitation to this approach, namely that the theory applied to an isotropically flexible polymer whose minimum-energy conformation in the linear state is that of a straight rod. By the early- to mid-1980s experiments showed that sequence-dependent conformational properties of DNA can contribute to deviations from the straight-rod minimum-energy state.[8,9] These observations motivated the development of Monte Carlo-based computational tools that could account for sequence-dependent effects on the cyclization probability.

Levene and Crothers[10,11] used Monte Carlo simulation to compute the distributions of chain end-to-end distance, mutual orientations of chain termini, and an estimate of the writhe distribution in closed chains required to compute the cyclization probability.[12] These calculations were combined, under the assumption of independent twist and writhe variables, with an analytical treatment of the closed-chain twist distribution in order to arrive at a comprehensive treatment of the cyclization problem. The problem of sample attrition, which is significant in the case of limited Monte Carlo ensembles of chain conformations, was dealt with by using a combinatorial chain-dimerization method, first explored by Alexandrowicz.[13] Around the same time, Hagerman,[14] and later Ramadevi and Hagerman,[15] published a Monte Carlo method that was based mainly on the contributions to the cyclization probability derived from the chain's axial degrees of freedom.

Motivated by the discovery of sequence-dependent intrinsic bends in DNA,[8] the Monte Carlo method showed that an intrinsic DNA bend comparable in magnitude to one that was experimentally characterized[16] can affect the value of the torsion-independent cyclization probability by fifty fold or more.[10] This result suggested that cyclization measurements are a powerful method for measuring the extent of intrinsic DNA bending due either to DNA-sequence-dependent conformational preferences or bending distortion induced by site-specific protein binding. Measurements of DNA bending by ligase-catalyzed cyclization was used with great effectiveness by Don Crothers and numerous coworkers during the last two decades of his career.[17-23]

As with most simulation-based methods, computing cyclization probabilities by Monte Carlo techniques involves a trade-off between speed and accuracy. The large amounts of computer time required to calculate cyclization probabilities for any specified DNA conformation limited the theory's usefulness as a tool for fitting experimental data. This limitation led Zhang and Crothers to develop a new statistical-mechanical approach in the early 2000s that was based on approximating the chain-conformation distribution in terms of harmonic fluctuations about the chain's mechanical-energy minimum.[24] Termed the harmonic approximation (HA), this method was the natural successor to the Shimada-Yamakawa theory, but had the virtue that calculations could be done for inhomogeneous DNA conformations having non-zero values of the helical parameters tilt and roll, or sequence-dependent elastic-energy constants. The harmonic approximation was subsequently extended to the problem of protein-mediated DNA looping, wherein the protein assembly mediating the loop was treated as a connected set of rigid subunits interacting through the same harmonic expression that governs interactions between base pairs in the DNA loop.[25] The entire looped structure, DNA and associated protein subunits, was treated as a circular polymer with the virtual chain segments defining protein-DNA and protein-protein contacts assumed to adopt non-canonical helical parameters. HA calculations are about four orders of magnitude faster than Monte Carlo simulations of the same system, which made it possible to carry out systematic analyses of lac-repressor-mediated DNA looping *in vitro*[26] and *in vivo*.[27]

The applicability of HA is limited to systems for which the accessible conformation space is dominated by harmonic fluctuations about a well-defined elastic-energy minimum. This assumption is justified, *e.g.*, for biopolymers with $L \ll P$ where $L$ is the contour length and $P$ is the persistence length. However, for larger systems HA results are expected to deviate from exact behavior because for increasing ratios $L/P$ the accessible conformation space is increasingly dominated by anharmonic contributions. For example, in the limit of long, flexible polymer chains the entropy of a circular chain of contour length $L$ *decreases* relative to the entropy of a linear chain of same length as $-3/2 \ln(L)$;[28,29] this logarithmic decrease as a function of $L$ is absent in the HA result for the entropy of this system. The accuracy and applicability of HA as a function of system size, for example the contour length $L$ of a cyclized DNA or a DNA loop, have not been systematically investigated. HA therefore remains an uncontrolled approximation for most systems.

The objective of the present study is to test the validity of HA for the cyclization probability or J factor of a simple homogeneous wormlike chain without torsional elastic energy. To this end, we compare values of the J factor computed by the method of normal-mode analysis (NMA), which is rigorously equivalent to HA computations,[24] with the corresponding exact result for the J factor computed by a new method that combines NMA with thermodynamic integration, a technique we denote TI-NMA.[30] This allows us to characterize the validity of HA (and NMA) in terms of a universal, model-independent function which depends only on the ratio $L/P$, where $L$ is the contour length and $P$ is the persistence length of the semiflexible chain. Although we consider here the simple case of a homogeneous wormlike chain without torsional elastic energy, we argue that our result for the deviation of the HA (and NMA) from the exact behavior as a function of $L/P$ qualitatively holds for any semiflexible macromolecular system which can be characterized by a contour length $L$ and a persistence length $P$, including helical wormlike chains, looped DNA, and DNA having intrinsic bends or other local flexible defects.

The rationale of this hypothesis is as follows. As discussed above, the validity of the HA (and NMA) is based on the assumption that the system undergoes harmonic fluctuations about a well-defined elastic-energy minimum. For increasing system size, *e.g.*, increasing contour length $L$ of a semiflexible biopolymer, it is the *large-scale conformational fluctuations* of the system, *i.e.*, those occurring on length scales large compared with the persistence length $P$, that generate the anharmonic contributions to the accessible conformation space and result in deviations of HA (and NMA) from the exact behavior. The precise form of the conformational fluctuations, and thus the free energy, of a complex macromolecular system depends of course on details such as monomer composition and solution conditions; however, the length scale on which these conformational fluctuations result in anharmonic behavior is set, by definition, by the persistence length $P$. This implies that the range of validity of HA and NMA is controlled by a single parameter, namely the ratio of system size to persistence length, or $L/P$. Due to the universal nature of our results for the deviation of HA (and NMA) from the exact behavior, as a function of $L/P$, we expect our results to remain qualitatively valid for more complex macromolecular systems.

## II. THEORY AND RESULTS

### 1. The cyclization probability, or J factor

Cyclization probabilities are expressed in terms of a thermodynamic quantity *J*, also called the J factor. *J* is defined as a ratio of equilibrium constants for intra- and intermolecular synapsis reactions of chains with $N$ vertices (or monomers) (Fig. 1),

$$J(N) = \frac{K_c(N)}{K_b} \quad . \tag{1}$$

In this expression,

$$K_c(N) = \frac{Q_{cir}(N)}{Q_{lin}(N)} \tag{2}$$

is the equilibrium constant for an intramolecular cyclization reaction in which the two vertices at the ends of a linear chain with $N$ vertices associate to form a circular chain with $N$ vertices. $Q_{cir}(N)$ and $Q_{lin}(N)$ are conformational partition functions of a circular chain (cir) and a linear chain (lin) with $N$ vertices, respectively. Partition functions are unitless by definition, which implies that the equilibrium constant $K_c(N)$ in Eq. (2) is unitless. In Eq. (1), the quantity

$$K_b = \frac{VQ_{lin}(2N)}{Q_{lin}(N) \cdot Q_{lin}(N)} \tag{3}$$

is the equilibrium constant for an intermolecular reaction in which two vertices at the ends of two different linear chains with $N$ vertices associate to form a linear chain with $2N$ vertices. The equilibrium constant $K_b$ is independent of $N$ because the dependence of $Q_{lin}$ on $N$ cancels in Eq. (3), reflecting the fact that for linear chains the conformational degrees of freedom of vertices are independent from each other. Only the enthalpy of the additional bond in the $2N$-chain enters $K_b$. Because the reaction corresponding to Eq. (3) is bimolecular, the quotient of partition functions in Eq. (3) is proportional to $c = 1/V$ where $c$ is the concentration of linear chains and $V$ is the available volume per chain.[31] The dependence on the concentration $c = 1/V$ is eliminated by the factor $V$ in the numerator of Eq. (3). This implies that $K_b$ has units of volume, the J factor in Eq. (1) has units of concentration, and both $K_b$ and $J$ are independent of $c = 1/V$. The J factor may be defined as the concentration of one end vertex in the vicinity of the other end vertex of the same linear chain L in the cyclization reaction shown in Fig. 1.[32,33]

The enthalpy of interaction between chain termini contributes similarly to $K_c(N)$ and $K_b$, which implies that the J factor is independent of the interaction enthalpy and depends only on elastic properties of the chain (the cancelation of the interaction enthalpy between chain termini is based on the well-supported assumption that the equilibrium constants for cyclization and bimolecular ligation depend on the details of the joining reaction in the same way). Therefore, testing the validity of NMA using the J factor yields a universal, model-independent result for the validity of NMA which depends only on the ratio $L/P$, where $L$ is the contour length and $P$ is the persistence length of the chain (cf. Section 4). Conversely, the free energy of cyclization

$$\Delta F = F_{cir} - F_{lin} = -k_B T \ln\left(\frac{Q_{cir}}{Q_{lin}}\right) \tag{4}$$

associated with the reaction $L \rightarrow C$ shown in Fig. 1 includes the association enthalpy and therefore depends on details of the interaction between chain termini.

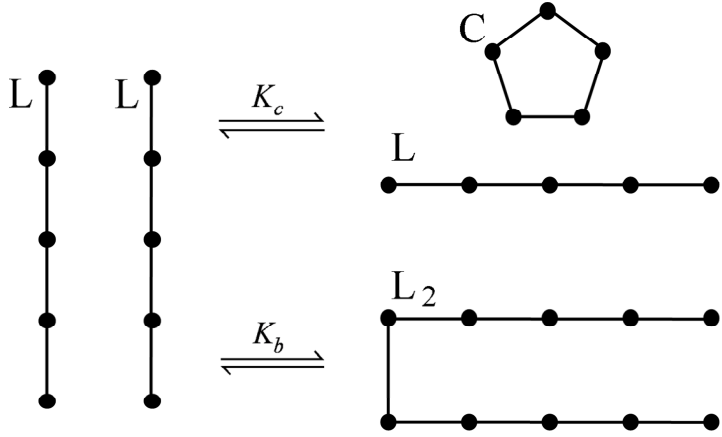

**Fig. 1**. Intra- and intermolecular synapsis reactions. For the intramolecular reaction (equilibrium constant $K_c$) the end vertices of a linear chain, L, with $N$ vertices and $N-1$ segments bind to form a circular chain, C, with $N$ vertices and $N$ segments (here $N=5$). The added segment in the circular chain corresponds to the new chemical bond formed in the cyclization reaction. For the intermolecular reaction (equilibrium constant $K_b$) the end vertices of two different linear chains L associate to form a linear chain with $2N$ vertices ($L_2$).

In what follows we calculate the equilibrium constants $K_c(N)$ and $K_b$ in Eq. (1) for a semi-flexible harmonic model chain. In Section 2 we derive an explicit expression for the partition function $Q_{lin}$ for a linear chain. We also obtain an explicit expression for the harmonic approximation $Q_{lin}^{(NMA)}$ of $Q_{lin}$, which is equivalent to calculating $Q_{lin}$ using normal-mode analysis (NMA). The equilibrium constants $K_b$ and $K_b^{(NMA)}$ are then calculated using Eq. (3). In Section 3 we discuss the numerical computation of the partition function $Q_{cir}$ for a circular chain using the TI-NMA method[30] and obtain the corresponding NMA result $Q_{cir}^{(NMA)}$. In Section 4 we obtain the J factor, $J$, as a universal function of $L/P$ and compare the exact result for $J$ with the corresponding NMA result, $J_{NMA}$. This allows us to assess the validity of NMA in terms of a universal function of $L/P$ which is free of microscopic details of our model chain. We argue that our results for the validity of NMA obtained here for a simple homogeneous wormlike chain characterize the validity of NMA for any semiflexible macromolecular system which can be characterized by a contour length $L$ and a persistence length $P$.

## 2. Partition function for a linear chain

We consider a semi-flexible harmonic chain as a coarse-grained mesoscopic model for duplex DNA. Chain elements are extensible segments with equilibrium length $b_0$ connected end-to-end by semi-flexible joints, or vertices, at positions $\mathbf{r}_i$, $i = 1,\ldots,N$ (Fig. 2). The conformational partition function of a linear chain with $N$ vertices (including the 2 vertices at the chain ends) and $N-1$ segments suspended in a volume $V$ is given by

$$Q_{lin}(N) = \int_V \frac{d^3 r_1}{a^3} \ldots \int_V \frac{d^3 r_N}{a^3} \exp\left[-\beta U_{lin}(\vec{r})\right] \tag{5}$$

where $U_{lin}(\vec{r})$ is the total potential energy for a chain conformation $\vec{r} = (\mathbf{r}_1,\ldots,\mathbf{r}_N)$ and $\beta = (k_B T)^{-1}$ ($T$ is the temperature in Kelvin and $k_B$ is the Boltzmann constant). The constant $a$ in Eq. (5) is a microscopic length required to make the partition function unitless. For a system of massive point particles undergoing Newtonian dynamics, the length $a$ corresponds to the thermal wavelength;[34] however, in this work we are concerned only with conformational degrees of freedom, and consider $a$ as a non-universal microscopic length much shorter than any other length scale associated with the chain. Essentially, the length $a$ corresponds to the lattice constant of an underlying lattice needed to obtain a finite number of accessible conformations. The results for the J factor obtained in this work are independent of the length $a$, and thus largely independent of the discretization of our model chain.

Segments are described by displacement vectors $\mathbf{b}_i = \mathbf{r}_{i+1} - \mathbf{r}_i$ with length $b_i$ and unit-length direction vectors $\hat{\mathbf{b}}_i = \mathbf{b}_i / b_i$. The bending angle $\theta_i$ at vertex $\mathbf{r}_i$ between segments $\mathbf{b}_{i-1}$ and $\mathbf{b}_i$ is given by $\cos(\theta_i) = \hat{\mathbf{b}}_{i-1} \cdot \hat{\mathbf{b}}_i$ (Fig. 2). The total potential energy of the chain is given by

$$U_{lin}(\vec{r}) = k_B T c_b \sum_{i=2}^{N-1} \left[1 - \cos(\theta_i)\right] + k_B T \frac{c_s}{2} \sum_{i=1}^{N-1} \left(\frac{b_i}{b_0} - 1\right)^2 , \tag{6}$$

where $c_b$, $c_s$ are bending and stretching elastic constants, respectively. In this work we neglect excluded volume interactions between chain segments, so that $U_{lin}(\vec{r})$ only includes the elastic potential energy of the chain. For a linear chain there is no elastic energy of bending associated with the end vertices $\mathbf{r}_1$ and $\mathbf{r}_N$ which implies that the first sum in Eq. (6) only includes the $N-2$ inner vertices $i = 2,\ldots,N-1$ (Fig. 2).

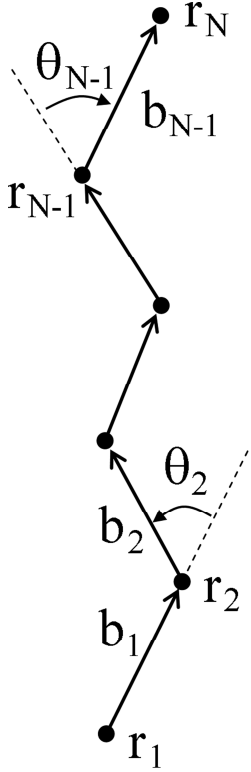

**Fig. 2.** Linear semi-flexible harmonic chain with $N$ vertices at positions $\vec{r} = (\mathbf{r}_1,\ldots,\mathbf{r}_N)$ and $N-1$ segments described by displacement vectors $\mathbf{b}_i = \mathbf{r}_{i+1} - \mathbf{r}_i$, $i = 1,\ldots,N-1$ (here $N=6$). Bending of the chain is described by $N-2$ polar angles $\theta_i$ located at inner vertices $\mathbf{r}_i$ between segments $\mathbf{b}_{i-1}$ and $\mathbf{b}_i$, $i = 2,\ldots,N-1$.

The bending energy constant $c_b$ in Eq. (6) is chosen such that the chain has a given persistence length $P$. Thus, $c_b$ is implicitly determined by the equation

$$\langle \cos(\theta) \rangle = \exp\left(-\frac{b_0}{P}\right) = \exp\left(-\frac{1}{n}\right), \tag{7}$$

where $n = P/b_0$ is the number of segments with equilibrium length $b_0$ per persistence length $P$. The thermal average $\langle \cos(\theta) \rangle$ is given by

$$\langle \cos(\theta) \rangle = \frac{\int_0^\pi d\theta \sin(\theta)\cos(\theta)\exp\left[-c_b(1-\cos(\theta))\right]}{\int_0^\pi d\theta \sin(\theta)\exp\left[-c_b(1-\cos(\theta))\right]}. \tag{8}$$

In this work we consider chains for which $n = \{150, 50, 25\}$. These $n$ values correspond to equilibrium segment lengths $b_0 = \{1, 3, 6\} \ell$ where $\ell$ is the DNA axial rise corresponding to a single base pair (using $P = 50$ nm for DNA under physiological conditions we obtain $\ell = P/150 = 0.3333$ nm). Numerically solving Eq. (7) for $c_b$ we find $c_b = \{150.5006, 50.5017, 25.5033\}$. The stretching energy constant $c_s$ in Eq. (6) is defined as $c_s = K_s b_0 / (k_B T)$ where $K_s$ is the stretching modulus. Using $b_0 = \{0.3333, 1, 2\}$ nm, $T = 300$ K, and the approximate value $K_s = 1000$ pN for DNA under physiological conditions, we find $c_s = \{80.4775, 241.4324, 482.8648\}$.

The reason for the above choice of the three $n$ values is that numerical simulations used to obtain exact results for the free energy of a circular chain (Section 3) are computationally most efficient for chains having a number of vertices $N$ between 50 and 300. For fixed $n$ this corresponds to a range of values $L/P = N/n$ spanning only a factor of 6 (e.g., $0.3333 \leq L/P \leq 2$ for $n = 150$). Given such a restricted range of $N$, choosing different values for $n$ allows us to vary $L/P$ over a much wider range, namely between 0.3333 ($N = 50$, $n = 150$) and 12 ($N = 300$, $n = 25$), corresponding to a 36-fold range of $L/P$. In addition, and equally important, using different $n$ values allows us to compute the J factor for the same value of $L/P$ using chains with different discretization (i.e., different segment lengths $b_0$ and associated elastic constants $c_b$, $c_s$). Because for a homogeneous chain we expect the J factor to depend only on the ratio $L/P$ on general grounds, we expect that the results for the J factor computed for chains with different $n$ collapse onto a single curve as a function of $L/P$. Using different $n$ values therefore provides not only an opportunity to test the expected universal behavior of the J factor for a homogeneous chain (namely its dependence on $L/P$ only, independent of details of the model chain), but also constitutes an important test of the validity and accuracy of our computational method.

The partition function $Q_{lin}(N)$ in Eq. (5) may be calculated explicitly by using spherical polar coordinates and carrying out the $N-2$ integrations over polar angles $\theta_i$ and $N-1$ radial integrations of segment lengths $b_i$ iteratively. We find

$$Q_{lin}(N) = \frac{V}{a^3}\left(\frac{b_0^3}{a^3}\right)^{N-1} 2(2\pi)^{3(N-1)/2}\left(\frac{1-e^{-2c_b}}{c_b}\right)^{N-2}\left(\frac{1+c_s}{c_s^{3/2}}\right)^{N-1}. \qquad (9)$$

To obtain Eq. (9) we used the approximation $\int_0^\infty db_i \simeq \int_{-\infty}^\infty db_i$, i.e., extending the lower integration limits for integrations of $b_i$ from 0 to $-\infty$; the error of this approximation is of order $\exp(-c_s/2)$ and is completely negligible for the values of $c_s$ used in this work. The harmonic approximation to $Q_{lin}(N)$, equivalent to calculating $Q_{lin}(N)$ by normal model analysis (NMA), may formally be obtained as the leading contribution of $Q_{lin}(N)$ in Eq. (9) in a Taylor expansion of $1/c_b$ and $1/c_s$ about $1/c_b = 0$ and $1/c_s = 0$, which is given by

$$Q_{lin}^{(NMA)}(N) = \frac{V}{a^3}\left(\frac{b_0^3}{a^3}\right)^{N-1} 2(2\pi)^{3(N-1)/2} c_b^{-(N-2)} c_s^{-(N-1)/2}. \qquad (10)$$

(This is formally equivalent to keeping for the exponent $\beta U_{lin}(\vec{r})$ of the Boltzmann factor in Eq. (5) only the leading, quadratic terms $\sim c_b \theta_i^2$ and $\sim c_s (b_i - b_0)^2$, and evaluating the resulting Gaussian integrals using $\int_0^\infty db_i b_i^2 \simeq \int_{-\infty}^\infty db_i b_i^2 \simeq b_0^2 \int_{-\infty}^\infty db_i$ in leading order.) The equilibrium constant $K_b$ is found by inserting $Q_{lin}$ from Eq. (9) in Eq. (3), resulting in

$$K_b = b_0^3 \, 2^{1/2} \pi^{3/2}\left(\frac{1-e^{-2c_b}}{c_b}\right)^2 \frac{1+c_s}{c_s^{3/2}}. \qquad (11)$$

Similarly, the NMA result $K_b^{(NMA)}$ is found by inserting $Q_{lin}^{(NMA)}$ from Eq. (10) in Eq. (3), resulting in

$$K_b^{(NMA)} = b_0^3 \, 2^{1/2} \pi^{3/2} c_b^{-2} c_s^{-1/2}. \qquad (12)$$

As noted above, $K_b$ has units of volume, but is independent of the concentration $c = 1/V$ of linear chains; in the present case, $K_b$ has units of volume because of the term $b_0^3$ in Eqs. (11), (12), where $b_0$ is the equilibrium length of a segment in our model chain.

## 3. Partition function and free energy for a circular chain

**3.1. TI-NMA method.** The partition function $Q_{cir}(N)$ for a circular chain with $N$ vertices and $N$ segments, corresponding to the circular state C shown in Fig. 1, is given by the expression in Eq. (5) replacing $U_{lin}(\vec{r})$ with

$$U_{el}(\vec{r}) = k_B T c_b \sum_{i=1}^{N}\left[1-\cos(\theta_i)\right] + k_B T \frac{c_s}{2}\sum_{i=1}^{N}\left(\frac{b_i}{b_0}-1\right)^2 \quad . \tag{13}$$

In this work, we neglect excluded volume interactions between chain segments and consider a phantom chain without topological constraint, i.e., we consider ensembles of circular chains which include all knot types. Furthermore, we consider the elastic energy due to bending and stretching of the chain only, as appropriate for nicked DNA. The only contribution to the potential energy of the chain is thus given by the elastic potential energy in Eq. (13) where the elastic constants $c_b$, $c_s$ are the same as in Section 2. We compute the partition function $Q_{cir}$ and the free energy $F_{cir} = -k_B T \ln(Q_{cir})$ using the TI-NMA method presented in reference [30] and summarized for the present case in Figure 3. In this method, a circular molecular state, $C$, is gradually transformed into a harmonically constrained reference state $C_0$ which corresponds to the minimum energy conformation. The associated change in free energy, $\Delta F^{(TI)}$, is computed by thermodynamic integration (TI). The absolute free energy, $F_0^{(NMA)}$, of the reference state $C_0$ is computed separately by using NMA. The TI-NMA method yields the absolute free energy of the circular state C as (Fig. 3)

$$F_{cir} = F_0^{(NMA)} - \Delta F^{(TI)} \quad . \tag{14}$$

**3.2. Normal-mode Analysis.** Applying normal-mode analysis (NMA) to the partition function $Q_{cir}$ yields an approximation, $Q_{cir}^{(NMA)}$, which is expressed in terms of the eigenvalues of the Hessian matrix associated with the potential function $U_{el}(\vec{r})$ in Eq. (13). The Hessian matrix is calculated at the minimum energy conformation $\vec{r}_0$ of the circular chain, which consists of a regular polygon with $N$ sides of length $b_0$ (Fig. 1). Following the procedure outlined in [30] we obtain

$$Q_{cir}^{(NMA)}(N) = \exp(-\beta E_0) \frac{V}{a^3} \left(\frac{b_0^3}{a^3}\right)^{N-1} N^{3/2} 8\pi^2 \sqrt{I_x I_y I_z} \prod_{m=7}^{3N} \left(\frac{2\pi}{v_m}\right)^{1/2} \tag{15}$$

where $E_0 = k_B T c_b N [1 - \cos(2\pi/N)]$ is the energy of the minimum conformation $\vec{r}_0$ and $I_x, I_y, I_z$ are the principal moments of inertia of $\vec{r}_0$ in units of $b_0$ (we here assume that each vertex of the chain is associated with a unit mass). The unitless quantities $v_m$ in Eq. (15) are the eigenvalues of the Hessian matrix in units of $k_B T$ and $b_0$. Assuming that the eigenvalues are ordered such that $v_1 \leq v_2 \leq \ldots \leq v_{3N}$ one finds $v_m = 0$ for $m = 1, \ldots, 6$ and $v_m > 0$ for $m = 7, \ldots, 3N$. The $3N - 6$ nonzero eigenvalues are associated with internal vibrations of the chain about the minimum conformation $\vec{r}_0$ which incur a finite energetic cost. Conversely, the 6 zero eigenvalues $v_m = 0$ for $m = 1, \ldots, 6$ are associated with rigid translations and rotations of the chain which do not incur any energetic cost. The corresponding eigenmodes contribute to $Q_{cir}^{(NMA)}(N)$ in Eq. (15) in terms of the number $N$ of "particles" (vertices) and on the shape of the energy-minimized conformation, $\vec{r}_0$, in terms of the principal moments $I_x, I_y, I_z$.[30]

**3.3. Thermodynamic integration.** The objective of thermodynamic integration (TI) is to gradually transform the original circular state C into a harmonically constrained reference state $C_0$ for which the corresponding free energy, $F_0^{(NMA)}$, may be computed accurately by applying NMA to $C_0$. To this end, we gradually replace the energy function $U_{el}$ of the original, semi-flexible circular chain C in Eq. (13) with a potential function corresponding to $C_0$ and to calculate the associated change in free energy, $\Delta F^{(TI)}$ (Fig. 3). A switching parameter $\lambda$ is used to effect this change in chain properties according to the following scheme

$$U(\lambda) = \begin{cases} \lambda U_{ha} + (1-\lambda) U_{el}, & 0 \leq \lambda \leq 1 \\ \lambda U_{ha}, & 1 \leq \lambda \leq \lambda_{max} \end{cases} . \tag{16}$$

The auxiliary elastic potential energy $U_{ha}$ serves to constrain the system to a predefined reference conformation, which is here given by the minimum energy conformation $\vec{r}_0$.[30] The auxiliary energy $U_{ha}$ is defined by implementing phased intrinsic bends at the vertices of the chain in such a way that for the given reference conformation $\vec{r}_0$ every chain segment points in its preferred direction relative to the preceding segment.

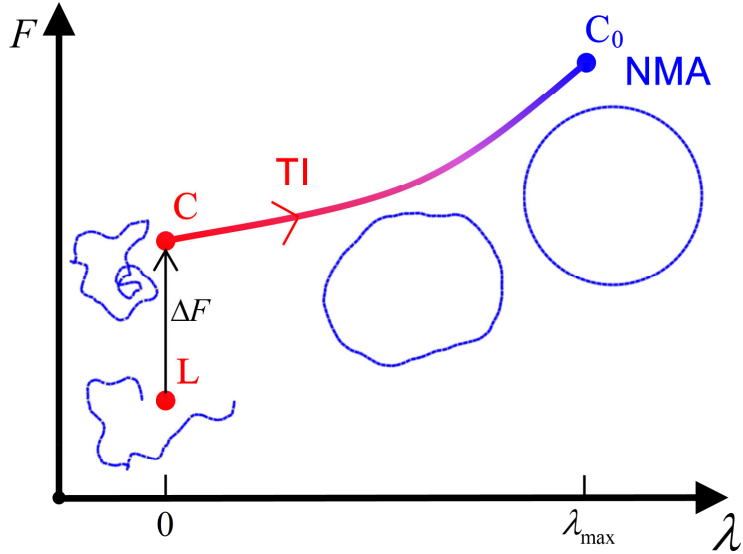

**Fig. 3**. Calculation of the cyclization free energy $\Delta F = F_{cir} - F_{lin}$ between a circular molecular state C and a linear molecular state L (cf. Fig. 1). Thermodynamic integration (TI) yields the change in free energy, $\Delta F^{(TI)}$, as the circular state C is transformed into the reference state $C_0$ (red section of the curve). Normal mode analysis (NMA) yields the absolute free energy $F_0^{(NMA)}$ of $C_0$ (blue section). The absolute free energy of the circular state C is then obtained as $F_{cir} = F_0^{(NMA)} - \Delta F^{(TI)}$. Typical Monte Carlo conformations of the circular state C are shown for the TI portion. The absolute free energy $F_{lin}$ of the linear state L is given by $F_{lin} = -k_B T \ln(Q_{lin})$ with $Q_{lin}$ given in Eq. (9).

TI is carried out in two phases, i.e., $\Delta F^{(TI)} = \Delta F_1^{(TI)} + \Delta F_2^{(TI)}$, with

$$\Delta F_1^{(TI)} = \int_0^1 d\lambda \left\langle \frac{dU}{d\lambda} \right\rangle_\lambda = \int_0^1 d\lambda \left\langle U_{ha} - U_{el} \right\rangle_\lambda, \tag{17}$$

$$\Delta F_2^{(TI)} = \int_1^{\lambda_{max}} d\lambda \left\langle \frac{dU}{d\lambda} \right\rangle_\lambda = \int_1^{\lambda_{max}} d\lambda \left\langle U_{ha} \right\rangle_\lambda. \tag{18}$$

Here $U = U(\lambda)$ is given by Eq. (16) and the symbol $\langle \ \rangle_\lambda$ indicates an ensemble average taken at a specific value of $\lambda$. Values for $\langle U_{ha} - U_{el} \rangle_\lambda$ in Eq. (17) and $\langle U_{ha} \rangle_\lambda$ in Eq. (18) were obtained by Monte Carlo simulation for 11 equally spaced values $\lambda = \{0, 0.1, 0.2, \ldots, 1.0\}$ and for exponentially increasing $\lambda$ values from 1 to a maximum value $\lambda_{max}$ (see [30] for details of the Monte Carlo simulation procedure). Starting with $\lambda = 1.0$, the values of $\lambda$ were increased according to $\lambda_{i+1} = 1.05 \lambda_i$ until $\lambda$ was large enough to satisfy the criterion $\langle U \rangle_\lambda = 0.5 \cdot (3N - 6) k_B T$ which holds in the harmonic regime due to the equipartition theorem (Fig. 4). The results were linearly interpolated and integrated according to Eqs. (17), (18). Each simulation was started at the minimum energy conformation $\bar{r}_0$ and an initial $1 \times 10^6$ trial moves were made to equilibrate the system. After initial equilibration, a new conformation was saved after each 1000 trial moves to produce a final ensemble. Fig. 4 shows $\langle U_{ha} - U_{el} \rangle_\lambda$ for $0 \leq \lambda \leq 1$ and $\langle \lambda U_{ha} \rangle_\lambda = \langle U \rangle_\lambda$ for $1 \leq \lambda \leq \lambda_{max}$ for all values of $\lambda$ quoted above. Each data point represents the average obtained from an ensemble of $10^5$ conformations.

### 4. Comparison of J-factor values obtained by NMA and TI-NMA

Using Eq. (1) one may express the J factor as

$$-\ln\left(\frac{J(N)}{c_0}\right) = \frac{\Delta F(N)}{k_B T} + \ln(c_0 K_b) \tag{19}$$

where the free energy of cyclization $\Delta F = F_{cir} - F_{lin}$ is calculated using Eqs. (4), (9), (14). In Eq. (14), the absolute free energy $F_0^{(NMA)}$ of the reference state $C_0$ is calculated using $F_0^{(NMA)} = -k_B T \ln(Q_{cir}^{(NMA)})$ with $Q_{cir}^{(NMA)}$ from Eq. (15) (cf. Fig. 3). Note that the common prefactor $\frac{V}{a^3} \left(\frac{b_0^3}{a^3}\right)^{N-1}$ in Eqs. (9) and (15) cancels in the difference $\Delta F = F_{cir} - F_{lin}$, which implies that the J factor is independent of the volume $V$ and of the microscopic length $a$ introduced in Eq. (5). The equilibrium constant $K_b$ in Eq. (19) is given by Eq. (11). The reference concentration $c_0$ in Eq. (19) is required to make the arguments of the logarithms in Eq. (19) unitless.

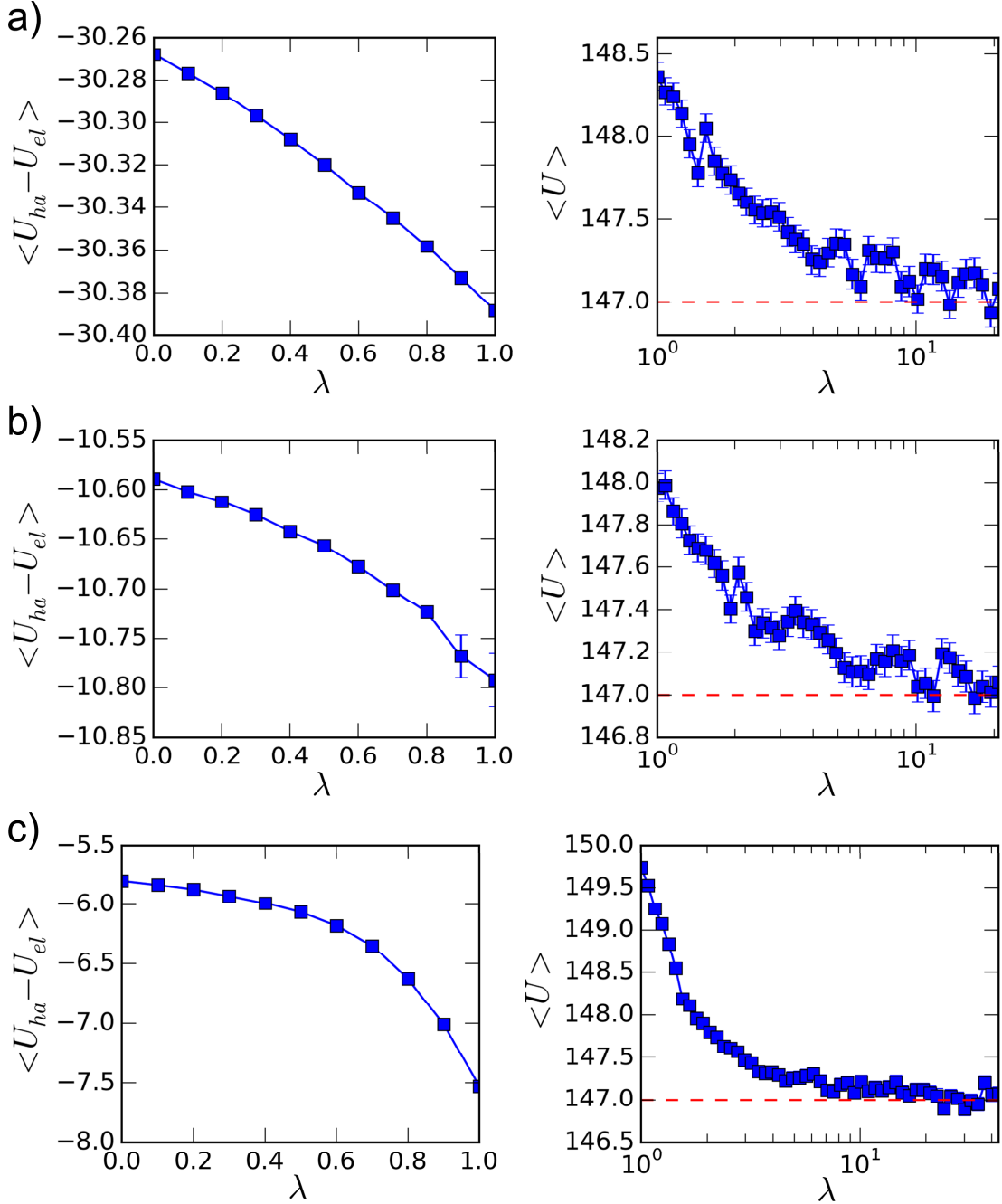

**Fig. 4**. $\langle U_{ha} - U_{el} \rangle_\lambda$ and $\langle \lambda U_{ha} \rangle_\lambda = \langle U \rangle_\lambda$ in units of $k_B T$ for all $\lambda$ values given in the text, for chains with $N=100$ segments and equilibrium segment lengths a) $b_0 = \ell$, b) $b_0 = 3\ell$, and c) $b_0 = 6\ell$ where $\ell$ is the axial rise per base pair in duplex DNA. In the harmonic regime obtained for large $\lambda$, $\langle U \rangle_\lambda / (k_B T)$ converges to $0.5 \cdot (3N-6) = 147$ for $N=100$ due to the equipartition theorem (dashed lines).

Note that $c_0$ merely defines the units scale in terms of which $J$ and $K_b$ are expressed; the quantities $J$ and $K_b$ themselves are independent of the concentrations $c_0$ and $c = 1/V$. In this work we express $J$ in units of the standard molar concentration $c_0 = 1$ M. Eq. (19) shows that the negative logarithm of the length-dependent J-factor $J(N)$ is a measure of the length-dependent cyclization free energy $\Delta F(N)$ modulo an additive constant. Both terms on the right hand side of Eq. (19) are non-universal, i.e., depend on details of the model, but $J$ on the left hand side of Eq. (19) is universal.

Similarly, the NMA result for the J factor is given by

$$-\ln\left(\frac{J_{NMA}}{c_0}\right) = \frac{\Delta F_{NMA}}{k_B T} + \ln\left(c_0 K_b^{(NMA)}\right) \quad , \tag{20}$$

where the NMA result for the free energy of cyclization, $\Delta F_{NMA}$, is obtained using $\Delta F_{NMA} = -k_B T \ln\left(Q_{cir}^{(NMA)} / Q_{lin}^{(NMA)}\right)$ with $Q_{lin}^{(NMA)}$ in Eq. (10). The NMA result $Q_{cir}^{(NMA)}$ is obtained by applying NMA directly to the circular state C (see Eq. (15) and Fig. 3). The equilibrium constant $K_b^{(NMA)}$ in Eq. (20) is given by Eq. (12).

$J$ and $J_{NMA}$ in Eqs. (19), (20) were calculated for chains with $N = \{50, 75, 100, 125, 150, 175, 200, 225, 250, 275, 300\}$ vertices and segments of equilibrium length $b_0 = \{1, 3, 6\}\ell$, where $\ell$ is the DNA axial rise corresponding to a single base pair (Fig. 5). The J factor, $J$, is universal in the sense that $J$ depends only on the ratio $L/P$, but not on microscopic details of the model, such as the segment length $b_0$. That is, when calculated using different values of $b_0$, the resulting J factor collapses to a single function of $L/P$. This collapse confirms the expected universal behavior of the J factor for a homogeneous chain and constitutes an acid test of the validity and accuracy of our computational method (Fig. 5).

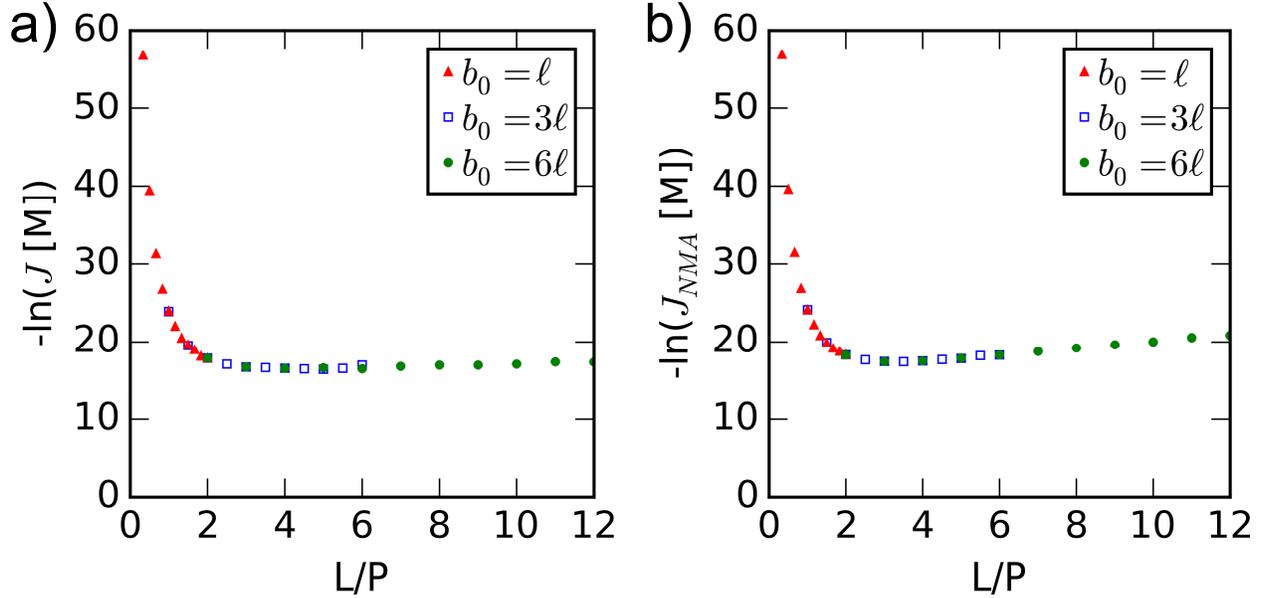

**Fig. 5.** a) $-\ln(J[M])$ and b) $-\ln(J_{NMA}[M])$ as universal functions of $L/P$, where $L$ is the contour length and $P$ the persistence length of the chains in the cyclization reaction (cf. Fig. 1). Because the J factor, $J$, is a universal function of $L/P$, results obtained using chains with different values of the segment length $b_0$ (here $b_0 = \{1, 3, 6\} \ell$ where $\ell$ is the rise per base pair in duplex DNA) collapse onto a single curve as a function of $L/P$.

Fig. 6 shows the deviation $\ln(J[M]) - \ln(J_{NMA}[M]) = \ln(J/J_{NMA})$ as a universal function of $L/P$. Note that the deviation $\ln(J[M]) - \ln(J_{NMA}[M])$ essentially corresponds to the difference in cyclization free energies $\Delta F_{NMA} - \Delta F$ in units of $k_B T$, modulo an additive constant. The length dependence of this positive deviation is essentially linear in the ratio $L/P$ and exceeds 1 $k_B T$ only for $L/P \gtrsim 4$, approximately 600 bp (assuming that $P$ = 50 nm). Thus, using the harmonic approximation or NMA for systems with smaller chain lengths yields a slightly underestimated J factor. Given that these results are a universal function of $L/P$, they should hold both for DNA cyclization and the more general problem of DNA looping.

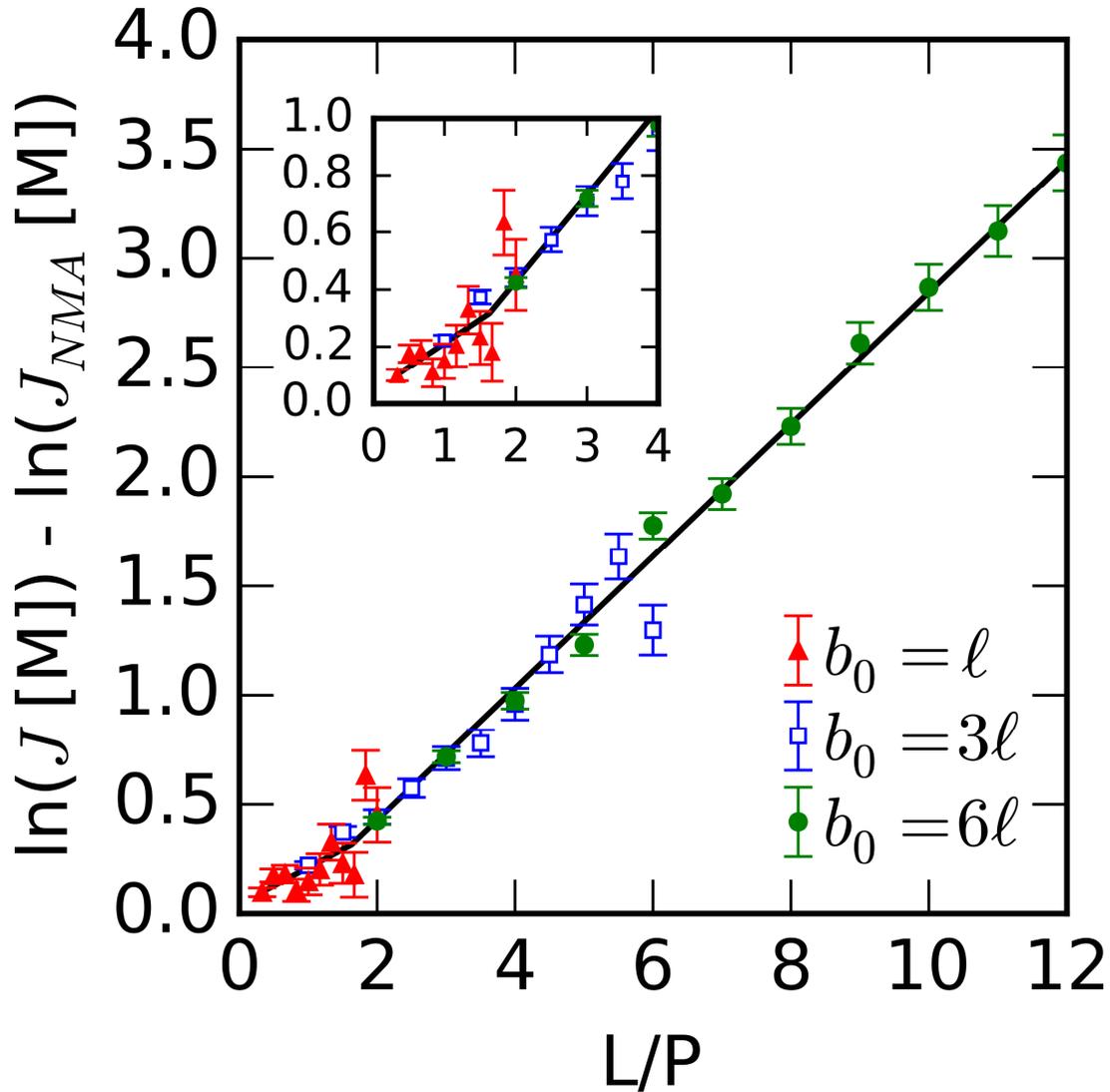

**Fig. 6.** Deviation $\ln(J[M]) - \ln(J_{NMA}[M])$ as a universal function of $L/P$ (cf. Fig. 5) to assess the validity of the harmonic approximation (HA) and normal mode analysis (NMA). The solid curve is a piecewise-linear fit given by $y = 0.166x + 0.0458$ ($x < 1.63$) and $y = 0.302x - 0.176$ ($x \geq 1.63$), where $x = L/P$ and $y = \ln(J[M]/J_{NMA}[M])$. The fit for $x \leq 4$ is also shown on an expanded scale in the inset to the figure. Since the deviation is obtained as a universal function of $L/P$, we expect similar behavior for any semiflexible macromolecular system which can be characterized by a contour length $L$ and persistence length $P$.

## III. CONCLUSIONS

There is increasing interest in the phenomenon of DNA and chromatin looping as a common mechanism of biological regulation.[35-44] Together with recognition that cyclization J-factor measurements are exquisitely sensitive to helical parameters and conformational properties of DNA molecules, there has been strong motivation to develop advanced statistical-mechanical models of DNA-loop formation (with cyclization as a special case). Although Monte Carlo methods for computing $J$ have been the standard for problems involving sequence-dependent bending and flexibility, they remain challenged by finite computing resources. HA and NMA-based J-factor calculations are up to four orders of magnitude more efficient and therefore more suitable for analyzing experimental looping and cyclization data. As attractive as these approaches are, they have been approximations of unknown extent to the actual physical behavior of DNA rings and loops.

We have sought here to estimate the error in the J factor computed by HA/NMA techniques, which arise from anharmonic contributions to the behavior of wormlike chains and increase with chain size. For $L/P \leq 4$, HA/NMA systematically underestimate the configurational free-energy cost of cyclization by an amount less than or equal to the thermal energy $k_B T$, as determined by rigorously computing the exact free energy using a thermodynamic integration technique. Whether this error is tolerable without a TI correction or not may depend on the accuracy needed for a given analysis. For J-factor measurements over a narrow range of DNA sizes, it may be sufficient to assume that a small, multiplicative factor greater than unity can be applied to $J_{NMA}$ in order to correct for the deviation. We suggest, however, that such an approach should be used with caution in the case of larger chains.

Although we consider here the simple test case of a homogeneous wormlike chain without torsional elastic energy, we argue that our result for the deviation of the HA (and NMA) from the exact behavior as a function of $L/P$ (Fig. 6) qualitatively holds for any semiflexible macromolecular system which can be characterized by a contour length $L$ and a persistence length $P$, including helical wormlike chains, looped DNA, and DNA having intrinsic bends or other locally flexible defects (cf. the end of Section I). It would be interesting to test this hypothesis by systematically testing the validity of HA and NMA for more complex systems.

A hallmark of Don Crothers' approach to science was his fearlessness in adapting, improving, or devising whatever mathematical or computational tools were needed to more effectively analyze experimental results. We offer the present theoretical treatment in the spirit of Don's legacy.

## IV. METHODS

Monte Carlo simulations were carried out using Fortran 90 and the same algorithm as in[30] but without monitoring excluded volume or knot checking. Averages $\langle U \rangle_\lambda$ and standard deviations $\sigma$ were obtained from ensembles of $10^5$ conformations each. In order to estimate the number of trial moves between conformations required to generate an ensemble of independent conformations, we calculated the autocorrelation function $\text{acf}(\tau) = \langle (U(t) - \langle U \rangle)(U(t+\tau) - \langle U \rangle) \rangle$ and fit $\text{acf}(\tau)$ to an exponential decay function $\sim \exp(-\tau/k)$ to estimate the value of the wait time $k$. Waiting times were measured in units of 1000 trial moves, *i.e.*, $k = 1$ corresponds to a waiting time of 1000 trial moves. If $k > 1/3$ was found, the number $\eta$ of independent conformations in the ensemble generated by the Monte Carlo simulation was estimated as $\eta = 10^5/(3k)$; if $k \leq 3$ was found, we used $\eta = 10^5$, *i.e.*, considered the entire ensemble of $10^5$ conformations as independent. Standard errors of the mean (sem) for each simulation was estimated as $\text{sem} = \sigma/\sqrt{\eta}$. Error in the TI procedure was estimated according to standard error-propagation analysis.

NMA calculations were performed using Python and Fortran 90. CPU time for this calculation scales as $\text{O}(N^2)$; for a chain of $N$ = 300 an NMA calculation takes under a minute on a single CPU. TI-NMA calculations were performed on a 32-CPU computing cluster. With all processors occupied a calculation for $N$ = 50 takes about one hour and scales linearly with large $N$ (*i.e.*, $N$ = 300 takes six hours). This linear scaling applies only to calculations omitting excluded-volume- and knot-type-checking steps. Source code is available upon request.

## ACKNOWLEDGEMENTS

This work was supported by grants from the NSF/NIGMS Initiative in Mathematical Biology (DMS-0800929) to SDL and NIH 2SC3GM083779 to AH.

**REFERENCES**

1. Shore, D.; Langowski, J.; Baldwin, R. L. Proc Natl Acad Sci USA 1981, 78, 4833-4837.
2. Shore, D.; Baldwin, R. L. J Mol Biol 1983, 170, 957-981.
3. Shore, D.; Baldwin, R. L. J Mol Biol 1983, 170, 983-1007.
4. Olson, W. K.; Marky, N. L.; Jernigan, R. L.; Zhurkin, V. B. J Mol Biol 1993, 232, 530-554.
5. Olson, W. K.; Zhurkin, V. B. Curr Opinion Struct Biol 2000, 10, 286-297.
6. Shimada, J.; Yamakawa, H. Macromolecules 1984, 17, 689-698.
7. Shimada, J.; Yamakawa, H. J Mol Biol 1985, 184, 319-329.
8. Marini, J. C.; Levene, S. D.; Crothers, D. M.; Englund, P. T. Proc Natl Acad Sci USA 1982, 79, 7664-7668.
9. Wu, H. M.; Crothers, D. M. Nature 1984, 308, 509-513.
10. Levene, S. D.; Crothers, D. M. J Mol Biol 1986, 189, 61-72.
11. Levene, S. D.; Crothers, D. M. J Mol Biol 1986, 189, 73-83.
12. Flory, P. J.; Suter, U. W.; Mutter, M. J Am Chem Soc 1976, 98, 5733-5739.
13. Alexandrowicz, Z. J Chem Phys 1969, 51, 561-565.
14. Hagerman, P. J. Biopolymers 1985, 24, 1881-1897.
15. Hagerman, P. J.; Ramadevi, V. A. J Mol Biol 1990, 212, 351-362.
16. Levene, S. D.; Wu, H. M.; Crothers, D. M. Biochemistry 1986, 25, 3988-3995.
17. Hockings, S. C.; Kahn, J. D.; Crothers, D. M. Proc Natl Acad Sci USA 1998, 95, 1410-1415.
18. Kahn, J. D.; Crothers, D. M. J Mol Biol 1998, 276, 287-309.
19. Kahn, J. D.; Yun, E.; Crothers, D. M. Nature 1994, 368, 163-166.
20. Koo, H. S.; Drak, J.; Rice, J. A.; Crothers, D. M. Biochemistry 1990, 29, 4227-4234.
21. Nathan, D.; Crothers, D. M. J Mol Biol 2002, 316, 7-17.
22. Roychoudhury, M.; Sitlani, A.; Lapham, J.; Crothers, D. M. Proc Natl Acad Sci USA 2000, 97, 13608-13613.
23. Zhang, Y.; Crothers, D. M. Proc Natl Acad Sci USA 2003, 100, 3161-3166.
24. Zhang, Y.; Crothers, D. M. Biophys J 2003, 84, 136-153.
25. Zhang, Y.; McEwen, A. E.; Crothers, D. M.; Levene, S. D. Biophys J 2006, 90, 1903-1912.
26. Swigon, D.; Coleman, B. D.; Olson, W. K. Proc Natl Acad Sci USA 2006, 103, 9879-9884.
27. Zhang, Y.; McEwen, A. E.; Crothers, D. M.; Levene, S. D. PLoS ONE 2006, 1, e136.

28. Flory, P. J. Principles of Polymer Chemistry; Cornell University Press: Ithaca, NY, 1953.
29. de Gennes, P. G. Scaling Concepts in Polymer Physics; Cornell University Press: Ithaca, NY, 1979.
30. Giovan, S. M.; Scharein, R. G.; Hanke, A.; Levene, S. D. J Chem Phys 2014, 141, 174902.
31. Hill, T. L. Introduction to Statistical Thermodynamics; Addison-Wesley: Reading, MA, 1960.
32. Crothers, D. M.; Drak, J.; Kahn, J. D.; Levene, S. D. Methods Enzymol 1992, 212, 3-29.
33. Levene, S. D.; Giovan, S. M.; Hanke, A.; Shoura, M. J. Biochem Soc Trans 2013, 41, 513-518.
34. Reichl, L. E. A Modern Course in Statistical Physics; John Wiley & Sons: New York, NY, 1998.
35. Polikanov, Y. S.; Bondarenko, V. A.; Tchernaenko, V.; Jiang, Y. I.; Lutter, L. C.; Vologodskii, A.; Studitsky, V. M. Biophys J 2007, 93, 2726-2731.
36. Hou, C.; Dale, R.; Dean, A. Proc Natl Acad Sci USA 2010, 107, 3651-3656.
37. Tokuda, N.; Sasai, M.; Chikenji, G. Biophys J 2011, 100, 126-134.
38. Guo, Y.; Monahan, K.; Wu, H.; Gertz, J.; Varley, K. E.; Li, W.; Myers, R. M.; Maniatis, T.; Wu, Q. Proc Natl Acad Sci USA 2012, 109, 21081-21086.
39. Kulaeva, O. I.; Nizovtseva, E. V.; Polikanov, Y. S.; Ulianov, S. V.; Studitsky, V. M. Mol Cell Biol 2012, 32, 4892-4897.
40. Le May, N.; Fradin, D.; Iltis, I.; Bougnères, P.; Egly, J.-M. Mol Cell 2012, 47, 622-632.
41. Sanyal, A.; Lajoie, B. R.; Jain, G.; Dekker, J. Nature 2012, 489, 109-113.
42. Brackley, C. A.; Taylor, S.; Papantonis, A.; Cook, P. R.; Marenduzzo, D. Proc Natl Acad Sci USA 2013, 110, E3605-E3611.
43. Chaumeil, J.; Micsinai, M.; Ntziachristos, P.; Deriano, L.; Wang, J. M.; Ji, Y.; Nora, E. P.; Rodesch, M. J.; Jeddeloh, J. A.; Aifantis, I.; Kluger, Y.; Schatz, D. G.; Skok, J. A. Cell Rep 2013, 3, 359-370.
44. Hensel, Z.; Weng, X.; Lagda, A. C.; Xiao, J. PLoS Biol 2013, 11, e1001591.